\documentclass[a4paper,10pt]{article}
\usepackage[utf8]{inputenc}
\usepackage[T1]{fontenc}
\usepackage{amsmath,amssymb,amsthm}
\usepackage[english]{babel}
\usepackage{graphicx}
\usepackage{listings}
\usepackage{dsfont}
\usepackage{geometry}
\usepackage{fancyvrb}
\usepackage{hyperref}
\begin{document}

\title{MPAgenomics : An R package for multi-patients analysis of genomic markers}
\author{Quentin Grimonprez\,$^{1,}$\footnote{to whom correspondence should be addressed}~,  Alain Celisse\,$^{1,2}$, Meyling Cheok$^{3}$, Martin Figeac$^{4}$\\ and Guillemette Marot\,$^{1,5}$}

\maketitle

\begin{center}
{$^{1}$ Modal team, Inria Lille-Nord Europe, France\\
$^{2}$ Laboratoire Paul Painlev\'e, Universit\'e Lille 1, France\\
$^{3}$ Inserm, U837, Team 3, Cancer Research Institute of Lille, France\\
$^{4}$ Plate-forme de génomique fonctionnelle et structurale, IFR-114, Universit\'e Lille 2, France \\
$^{5}$ EA 2694, Universit\'e Lille 2, France }
\end{center}

\begin{abstract}

\textbf{Summary:}
\texttt{MPAgenomics}, standing for multi-patients analysis (MPA) of genomic markers, is an R-package 
devoted to: $(i)$ efficient segmentation, and $(ii)$ genomic marker selection from multi-patient copy number and SNP data profiles.
It provides wrappers from commonly used packages to facilitate their repeated (sometimes difficult) use, offering an easy-to-use pipeline for beginners in R. 

The segmentation of successive multiple profiles (finding losses and gains) is based on a new automatic choice of influential parameters since default ones were misleading in the original packages. 
Considering multiple profiles in the same time, \texttt{MPAgenomics} wraps efficient penalized regression methods to select relevant markers associated with a given response.

\textbf{Availability:}
The R-package \texttt{MPAgenomics} is available on R-forge at \href{http://r-forge.r-project.org/R/?group\_id=1658}{http://r-forge.r-project.org/R/?group\_id=1658}.

\textbf{Contact:} \href{quentin.grimonprez@inria.fr}{quentin.grimonprez@inria.fr}

\end{abstract}

\section{Introduction}

Analyzing data from genome-wide SNP arrays within R requires several packages, e.g. \texttt{aroma} for normalization of Affymetrix SNP6.0 arrays \cite{aroma}, \texttt{changepoint} for segmentation of copy number profiles \cite{pelt}, \texttt{cghcall} for labelling segments \cite{cghcall}, and \texttt{glmnet} for penalized regressions \cite{glmnet}. 
Each package performs a specific task in the whole analysis but none of them is related to the others. Output formats of given packages are often not compatible with input formats required by the other, making their use awkward for beginners in R.
One main contribution of the \texttt{MPAgenomics} R package is to aggregate these commonly used packages, providing wrappers to inter-relate them automatically.

At each step of the analysis a large amount of packages are available to perform normalization, segmentation or marker selection. A careful choice of only a few methods is required to provide an easy-to-use and efficient tool. For instance the \texttt{MPAgenomics} segmentation step is based
on \textsc{Pelt} \cite{pelt}, which has been proved to be the most reliable method to segment copy number profiles among 17 competitors \cite{compseg}.
Furthermore the \texttt{MPAgenomics} package improves on the native \textsc{Pelt} method by providing an automatic data-driven choice of its penalty parameter.

In this application note, we describe two different pipelines implemented in the R package \texttt{MPAgenomics}. Both of them perform the whole analysis from raw data to normalization, and then to either successive segmented profiles, or a list of genomic markers selected from all available profiles.

\section{MPAgenomics package}
	\label{sec:single}

One interest of \texttt{MPAgenomics} is to provide a simple automatic way to combine the elementary steps described in what follows. Each of them can however be used separately by more advanced users.

\subsection{Data normalization}
\label{subsec:norm}

The normalization process in \texttt{MPAgenomics} contains \emph{technical biaises correction}, and \emph{copy number and allele B fraction estimation}. 
Following \cite{tumorboost}, \emph{allele B fraction} refers to the proportion of the total signal coming from allele B. 
Normalization methods are available for Affymetrix arrays (GenomeWideSNP 5 \& 6, \dots). 
The estimation of the total copy number and allele B fraction is made by \textit{CRMAv2} \cite{CRMAv2} originally implemented in the \texttt{aroma} packages. 
For studies with matched normal-tumor samples, a better estimation is suggested for the allele B fraction of the tumoral sample with the TumorBoost method \cite{tumorboost}.

The use of \texttt{aroma} packages is difficult for neophytes due to the complex folder architecture it requires and the lack of internal documentation of the R package. 
\texttt{MPAgenomics} implements a wrapper to process all these normalization steps and build the folder architecture automatically. 
Furthermore different graphs such as the copy number signal can be saved in the working folder architecture for further visualization.

\subsection{Segmentation method}

Following \cite{compseg} the \textsc{Pelt} segmentation method \cite{pelt} is implemented in \texttt{MPAgenomics}. It relies on a penalty $\lambda\log(n)$ penalizing too many segments, with a profile of length $n$ and $\lambda$ a parameter to choose. 
We observed that using the default parameter $\lambda=1$ on a real dataset of 70 profiles \cite{citerealdataset} leads to over-segmented regions (too many segments). 
Since $\lambda$ is crucial, \texttt{MPAgenomics} suggests an automatic sample-specific choice of $\lambda$ (see Section~\ref{sec:lambda}). 

The implemented segmentation method is available both for copy number and allele B fraction profiles. 
The allele B fraction segmentation is only made from heterozygous SNPs. The resulting signal is centered and symmetrized around 0, which makes it similar to the usual copy number.

\subsection{Calling method}

From each segmented profile, the \textit{CGHcall} method \cite{cghcall} is run to label every segment in terms of $loss$, $normal$, and $gain$. 
Segmentation and labeled segments are available in two settings. One is aroma-based and exploits the folder architecture and the files generated along the process. 
The second does not depend on aroma. It is particularly relevant for more advanced users with their own normalized data. 

In the aroma-dependent function, segmentation and calling are performed with the same wrapper. The calling is run for each profile separately. Results are saved in \textit{.BED} format in the working folder architecture.

\subsection{Genomic marker selection}

The goal is to select genomic markers (e.g. SNPs or CNV) associated with a given response from all patient profiles simultaneously. 

For each individual $i$ ($1\leq i\leq I$),  $y_i$ denotes the response and $x_{i,p}$ the corresponding normalized value of copy number or allele B fraction signal at genomic position $p$ ($1\leq p\leq P$). 
By default normalization is done as in Section~\ref{subsec:norm} without between-array normalization. There is no need to perform segmentation and calling before the multi-patients analysis.

Due to the huge number of markers ($P \gg I$) \texttt{MPAgenomics} uses the \emph{lasso} \cite{lasso} regularization method to select very few ones. 
It consists in minimizing $\beta \in \mathbb{R}^P\mapsto g(\beta)$, where
\begin{align*}
g_{\rho}(\beta) = \sum\limits_{i=1}^I (y_i-(X\beta)_i)^2+\rho \sum\limits_{p=1}^P |\beta_p|  \enspace,
\end{align*}
with $(X\beta)_{i}=\sum_p x_{i,p}\beta_p$ and $\rho>0$ controlling the number of non-zero coordinates of $\beta$.
After minimization, non-zero coefficients $\beta_p$ correspond to influential positions to predict the response.

\texttt{MPAgenomics} drastically improves on ongoing packages in terms of computation time. With the linear regression model, it efficiently provides the exact solution by use of the new R package \texttt{HDPenReg}, which is an optimized implementation of the \textit{lars} algorithm specially dedicated to huge number of markers.
Logistic regression is also available with binary responses. \texttt{MPAgenomics} wraps the \texttt{glmnet} package \cite{glmnet} in the whole process. Unlike \texttt{HDPenReg} it does not provide the exact solution but is computationally very efficient.

With \texttt{glmnet} and  \texttt{HDPenReg}, the regularization parameter $\rho$ is chosen by $k$-folds cross-validation \cite{arlotcelisse}. The selected variables are the most relevant ones regarding to the response.

\subsection{MPAgenomics vignette}

The tutorial of the \texttt{MPAgenomics} R package is obtained by running the following commands in the R console:

\texttt{library(MPAgenomics)}

\texttt{vignette(``MPAgenomics'')}

An example explains how to quickly analyze data. More details on each step or wrapper are given to help advanced users to run each function separately.

\section{Sample-specific parameter in MPA}
	\label{sec:lambda}

First we detail the \emph{sample-specific} choice of the $\lambda$ parameter we propose in the \textsc{Pelt} method (segmentation). Then we illustrate its potential improvement upon 
a common parameter choice on real data from \cite{citerealdataset}.

\subsection{Proposed method}
\label{sec.sample.specific}

\begin{figure}[!h]
		\centering
		\includegraphics[width=6cm,height=5cm]{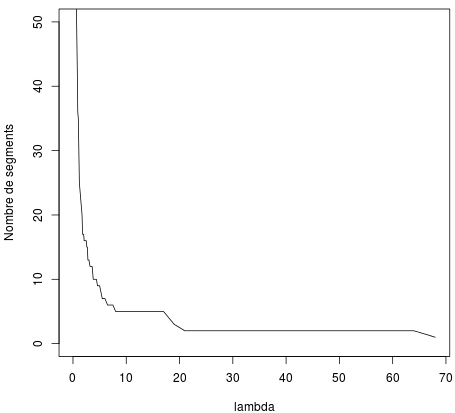}
		\caption{Number of segments for each $\lambda$ in the penalty of \textsc{Pelt}.}
		\label{fig:plateau}
	\end{figure}

For each profile the \textsc{Pelt} method is run on a grid of $\lambda$ values. 
Fig.~\ref{fig:plateau} displays the number of segments with respects to $\lambda$ on an example.
The widest range of $\lambda$ for which the number of segments remains unchanged (and larger than 1) indicates a high confidence in resulting segmentation.

The optimal data-driven $\lambda$ is the left-most value of the widest range such that the number of segments is larger than 1. Otherwise we consider there is only one segment in the profile.

\subsection{Sample-specific versus common parameter}

The \emph{sample-specific choice of $\lambda$} in Section~\ref{sec.sample.specific} is compared with a \emph{common choice of  $\lambda$} depending on the signal-to-noise ratio within each group of profiles.

\begin{figure}[!h]
	\centering
	\includegraphics[scale=0.3]{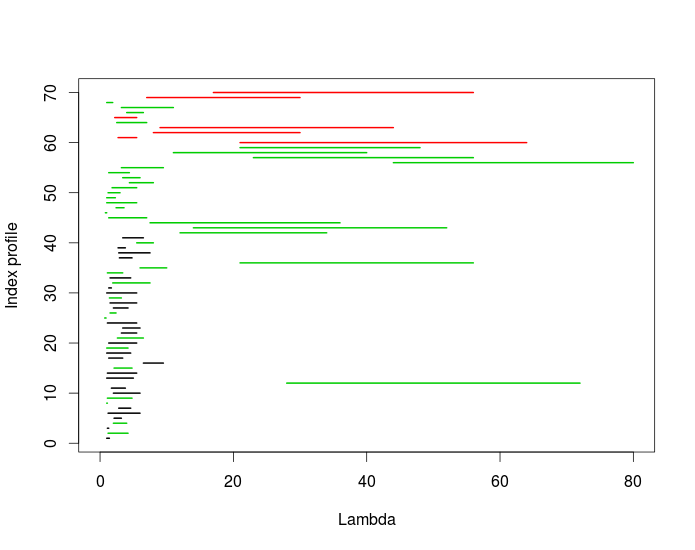}
	\caption{Widest ranges of $\lambda$ ($x$-axis) for 70 copy number profiles (chromosome 1) ($y$-axis). Colors indicate clusters of signal-to-noise ration (black $<$ red $<$ green).}
	\label{fig:classifplateau}
\end{figure}

Profiles from the real dataset \cite{citerealdataset} are clustered into groups with homogeneous signal-to-noise ratio (SNR) by use of Gaussian mixture model. Three groups are provided by the BIC criterion. 
Figure~\ref{fig:classifplateau} displays results (chromosome~1) for each profile (patient) from 1 to 70. Following Section~\ref{sec.sample.specific} the widest range of $\lambda$ values is plotted for each profile. 
Colors (black, red, and green) indicate the SNR level in each group (respectively low, middle, and high).

Whereas the lowest SNR group only contains ranges of $\lambda$ with small values, other groups correspond to ranges of both small and large values of $\lambda$.
A common choice of $\lambda$ within each of these two groups lead to erroneous segmentations.
The same conclusion applies to other chromosomes and criteria such as variance, which justifies the implementation of the sample-specific choice of $\lambda$ in \texttt{MPAgenomics}.

%%%------------------------------------------

\section{Conclusion}

\texttt{MPAgenomics} provides user-friendly wrappers for normalization and multi-patients analysis of genomic data. 
It also provides automatic choices of crucial parameters for segmentation and marker selection. 
Even though normalization is provided for Affymetrix arrays, others steps (segmentation, calling, and marker selection) can be applied to high-throughput sequencing data.

\section*{Acknowledgement}
We thank Serge Iovleff for his help implementing \texttt{HDPenReg}, and Samuel Blanck for relevant remarks when testing first versions of \texttt{MPAgenomics}. 
We also thank Claude Preudhomme and Olivier Nibourel for providing the data, and for their helpful clinical competences to interpret the results.  

\paragraph{Funding:} The developpment of this package was funded by the Inria Action de D\'eveloppement Technologique named \textit{MPAGenomics}.

\bibliographystyle{plain}

\bibliography{MPAgenomics_hal}

\begin{thebibliography}{10}

\bibitem{arlotcelisse}
Sylvain Arlot and Alain Celisse.
\newblock A survey of cross-validation procedures for model selection.
\newblock {\em Statistics Surveys}, 4:40--79, 2010.

\bibitem{aroma}
Henrik Bengtsson.
\newblock {aroma} - {A}n {R} {O}bject-oriented {M}icroarray {A}nalysis
  environment.
\newblock {Preprint in Mathematical Sciences} 2004:18, Mathematical Statistics,
  Centre for Mathematical Sciences, Lund University, Sweden, 2004.

\bibitem{tumorboost}
Henrik Bengtsson, Pierre Neuvial, and Terence~P. Speed.
\newblock Tumorboost: Normalization of allele-specific tumor copy numbers from
  a single pair of tumor-normal genotyping microarrays.
\newblock {\em BMC Bioinformatics}, 11, 2010.

\bibitem{CRMAv2}
Henrik Bengtsson, Pratyaska Wirapati, and Terence~P. Speed.
\newblock A single-array preprocessing method for estimating full-resolution
  raw copys from all affymetrix genotyping arrays including genomewidesnp 5 \&
  6.
\newblock {\em Bioinformatics}, 25(17):2149--2156, 2009.

\bibitem{glmnet}
Jerome~H. Friedman, Trevor Hastie, and Rob Tibshirani.
\newblock Regularization paths for generalized linear models via coordinate
  descent.
\newblock {\em Journal of Statistical Software}, 33(1):1--22, 2 2010.

\bibitem{compseg}
Toby Hocking, Gudrun Schleiermacher, Isabelle Janoueix-Lerosey, Valentina
  Boeva, Julie Cappo, Olivier Delattre, Francis Bach, and Jean-Philippe Vert.
\newblock Learning smoothing models of copy number profiles using breakpoint
  annotations.
\newblock {\em BMC Bioinformatics}, 14(1):164, 2013.

\bibitem{pelt}
Rebecca Killick and Idris Eckley.
\newblock {\em changepoint: An R package for changepoint analysis}, 2013.
\newblock R package version 1.1.

\bibitem{citerealdataset}
Aline Renneville, Raouf~Ben Abdelali, Sylvie Chevret, Olivier Nibourel, Meyling
  Cheok, Cecile Pautas, Remy Dulery, Thomas Boyer, Jean-Michel Cayuela,
  Sandrine Hayette, Emmanuel Raffoux, Hassan Farhat, Nicolas Boissel, Christine
  Terre, Herve Dombret, Sylvie Castaigne, and Claude Preudhomme.
\newblock Clinical impact of gene mutations and lesions detected by snp-array
  karyotyping in acute myeloid leukemia patients in the context of gemtuzumab
  ozogamicin treatment: Results of the alfa-0701 trial.
\newblock {\em Oncotarget}, 4(9), 2013.

\bibitem{lasso}
Robert Tibshirani.
\newblock Regression shrinkage and selection via the lasso.
\newblock {\em Journal of the Royal Statistical Society, Series B},
  58:267--288, 1994.

\bibitem{cghcall}
Mark van~de Wiel and Sjoerd Vosse.
\newblock {\em CGHcall: Calling aberrations for array CGH tumor profiles.},
  2012.
\newblock R package version 2.20.0.

\end{thebibliography}

\end{document}